\documentstyle[12pt]{article}
\newcommand{\gsim}{\mathrel{\hbox{\rlap{\lower.55ex \hbox {$\sim$}}
                   \kern-.3em \raise.4ex \hbox{$>$}}}}
\newcommand{\lsim}{\mathrel{\hbox{\rlap{\lower.55ex \hbox {$\sim$}}
                   \kern-.3em \raise.4ex \hbox{$<$}}}}
\newcommand {\vectwo}[2] {\left(\begin{array}{c}#1\\#2\end{array}\right)}

\newcommand {\beq}    {\begin{equation}}
\newcommand {\eeq}    {\end{equation}}

\newcommand {\thetaw} {\theta_{w}}
\newcommand {\dstar}  {d^{\star }}
\newcommand {\ES}     {E_{{\rm S}}}
\newcommand {\twoi}   {2\, i}
\newcommand {\alphatz}{\tilde{\alpha}_{0} }
\newcommand {\alphatr}{\tilde{\alpha}_{1} }

\newcommand {\ncl} {non-contractible loop}

\newcommand {\ncls}{non-contractible loops}
\newcommand {\ncs} {non-contractible sphere}
\newcommand {\Ncs} {Non-contractible sphere}
\newcommand {\ncss}{non-contractible spheres}
\newcommand {\bcs} {boundary conditions}
\newcommand {\sm}  {standard model}

\newcommand {\YMHth} {Yang-Mills-Higgs theory}

\newcommand {\sph}   {spha\-le\-ron}
\newcommand {\sphs}  {sphalerons}

\newcommand {\cs} {configuration space}
\newcommand {\ew} {electroweak}

\newcommand {\Sstar} {S$^{\star }$}
\newcommand {\Istar} {I$^{\star }$}

\newcommand {\scansatz} {self-consistent ansatz}
\newcommand {\ewsmodel} {electroweak standard model}

\begin{document}

\begin{titlepage}
\hspace*{\fill}
NIKHEF-H/93-09
\newline
\hspace*{\fill}
June, 1993
\begin{center}
	\vspace{2\baselineskip}
	{\Large \bf Construction of a new \ew ~sphaleron}\\
	\vspace{1\baselineskip}
\renewcommand{\thefootnote}{\fnsymbol{footnote}}
	{\large
	 F. R. Klinkhamer
\footnote{ ~E-mail address : klinkhamer @ nikhef.nl} \\
	 CHEAF/NIKHEF--H \\ Postbus 41882 \\
	 1009 DB Amsterdam \\ The Netherlands \\
	}
\renewcommand{\thefootnote}{\arabic{footnote}}
\setcounter{footnote}{0}
	\vspace{3\baselineskip}
	{\bf Abstract} \\
\end{center}
{\small \par
We present a self-consistent ansatz for a new \sph ~in the
\ew ~\sm. The resulting field equations are solved numerically.
This \sph ~sets the height of the energy barrier for the
global $SU(2)$ anomaly.
}
\end{titlepage}
%\end{document}

\section{Introduction}

In a series of papers \cite{K85,K90,K93} we have argued for the existence
of a new static, but unstable, classical solution \Sstar ~in the \ewsmodel.
The existence of this \sph ~\Sstar ~would be due to the presence
of \ncs s in \cs.
Here  we give the explicit construction of this new classical
solution. The self-consistent ansatz for \Sstar ~turns out to be
a direct generalization of the configuration at the top of the \ncs
{}~as constructed in our previous papers.
The same method can probably be used to discover other solutions,
as will be explained later on.

The outline of this article is as follows. In sect. 2 we present
the ansatz for the \sph ~\Sstar. The resulting expressions for
the energy and field equations are discussed in sect. 3.
The field equations are then solved numerically and some preliminary
results are given in sect. 4. Section 5 contains further remarks on the
\sph ~\Sstar ~and its physical interpretation.

\section{Ansatz}

We consider static classical fields in the bosonic sector of the \ewsmodel.
These are the $SU(2)$ gauge fields
$W_{m} \equiv W_{m}^{a}\: \sigma^{a} / (\twoi) \,$, the $U(1)$ hypercharge
gauge field $B_{m}$ and the complex Higgs doublet $\Phi$.
The $SU(2) \times U(1)$ \YMHth ~has an energy functional
\beq
 E = \int_{{\rm R}^3} d^{3}x
 \left[ \frac{1}{4 \,g^{2}}\; \left( W_{m n}^{a} \right)^{2} +
        \frac{1}{4 \,g^{\prime \,2}}\; \left( B_{m n}     \right)^{2} +
        |D_{m} \Phi|^{2} +
         \lambda \left( |\Phi|^{2} - v^{2}/2 \right)^{2} \: \right]
\label{eq:E}
\eeq
and field equations
\begin{eqnarray}
D_{m} W_{mn}   &=& g^{2} \: \left[
                   (D_{n} \Phi)^{\dagger}\;\frac{\sigma^{a}}{\twoi }\;\Phi -
                    \Phi^{\dagger}\;\frac{\sigma^{a}}{\twoi }\;(D_{n} \Phi)\:
                        \:  \right] \; \frac{\sigma^{a}}{\twoi }
                   \nonumber \\
\partial_{m} B_{mn}&=& g^{\prime \, 2} \left[\:
                         (D_{n} \Phi)^{\dagger}\;\frac{1}{\twoi }\;\Phi -
                          \Phi^{\dagger}\;\frac{1}{\twoi }\;(D_{n} \Phi)\;
                         \right]
                   \nonumber \\
D_{m}D_{m} \Phi    &=& 2\: \lambda
           \left[\Phi^{\dagger}\Phi -v^{2}/2\right]\:\Phi\; ,
\label{eq:fieldeqs}
\end{eqnarray}
with the following definitions for the field strengths and
covariant derivatives
\begin{eqnarray*}
W_{mn} &\equiv& W_{mn}^{a}\: \frac{\sigma^{a}}{\twoi}  \equiv
  \partial _{m} W_{n} - \partial _{n} W_{m} + [W_{m},W_{n}]\\
B_{m n} &\equiv&  \partial _{m} B_{n} - \partial _{n} B_{m}\\
D_{k}W_{lm} &\equiv&  \partial _{k} W_{lm} + [W_{k},W_{lm}] \\
D_{m}\Phi&\equiv&\left( \partial_{m} + W_{m}^{a}\; \frac{\sigma^{a}}{\twoi }
                         + B_{m}\; \frac{1}{\twoi } \right) \Phi \: ,
\end{eqnarray*}
where the indices run over the values $1,\, 2,\,3$, and
$\sigma^{a}$ are the standard Pauli matrices.
The semiclassical masses of the $W^{\pm}$ and $Z^{0}$ vector bosons are
$M_{W}=\frac{1}{2}\, g\, v$
and $M_{Z} = M_{W} / \cos \thetaw$, with the weak mixing angle
$\thetaw$ defined as $\tan \thetaw \equiv g' / g$.
The mass of the single Higgs scalar is $M_{H}=\sqrt{8\,\lambda/g^{2}}\, M_{W}$.
The classical \YMHth ~depends only on the two coupling constants
$\lambda/g^{2}$ and $\thetaw$.

Before we present the ansatz we introduce some notation.
The standard cylindrical coordinates $\rho$, $z$ and $\phi$  and spherical
coordinates $r$, $\theta$ and $\phi$ are defined
in terms of the cartesian coordinates by
\[
(x_{1},x_{2},x_{3}) \equiv (\rho \cos \phi\,, \rho \sin \phi\,,z)
\equiv (r \sin \theta \cos \phi\,, r \sin \theta \sin \phi\,,r \cos \theta) \:.
%\label{eq:coordinates}
\]
We also define three matrices
\newcommand {\matu}      {\underline{u}}
\newcommand {\matv}      {\underline{v}}
\newcommand {\matw}      {\underline{w}}
\begin{eqnarray*}
\matu&\equiv&
  +\cos\phi\;\frac{\sigma^{1}}{\twoi }+\sin\phi\;\frac{\sigma^{2}}{\twoi } \\
\matv&\equiv&
  -\sin\phi\;\frac{\sigma^{1}}{\twoi }+\cos\phi\;\frac{\sigma^{2}}{\twoi } \\
\matw&\equiv&
  \frac{\sigma^{3}}{\twoi } \: ,
%\label{eq:uvw}
\end{eqnarray*}
which take values in the Lie algebra of $SU(2)$.
These matrices have several useful properties \cite{M78}, such as
$\matu \cdot \matv = - \matv \cdot \matu = \frac{1}{2} \,\matw$
and further on cyclically.

We are now ready to present the ansatz.
The $SU(2)$ and $U(1)$ gauge fields and the Higgs doublet are given by
\begin{eqnarray}
W_{1} &=&  \frac{\alpha_{1}}{\rho}\cos\phi \; \matv
          +\frac{\alpha_{2}}{\rho}\sin\phi \; \matu
          +\frac{\alpha_{3}}{\rho}\sin\phi \; \matw \nonumber\\
W_{2} &=&  \frac{\alpha_{1}}{\rho}\sin\phi \; \matv
          -\frac{\alpha_{2}}{\rho}\cos\phi \; \matu
          -\frac{\alpha_{3}}{\rho}\cos\phi \; \matw \nonumber\\
W_{3} &=&  \frac{\alpha_{0}}{z}            \; \matv \nonumber\\
B_{1} &=&  \, \tan^{2}\thetaw \;\frac{\alpha_{4}}{\rho}\;\sin\phi \nonumber\\
B_{2} &=&  -\,\tan^{2}\thetaw \;\frac{\alpha_{4}}{\rho}\;\cos\phi \nonumber\\
B_{3} &=&  \, 0                                                   \nonumber\\
\Phi  &=& \left[ \,\beta_{1}\,(\cos\phi\;i\sigma^{1}+\sin\phi\; i\sigma^{2})
                -\beta_{2} \; i\sigma^{3}\,
          \right] \frac{v}{\sqrt{2}} \vectwo{0}{1}  \:,
\label{eq:ansatzW}
\end{eqnarray}
with
\begin{eqnarray}
\alpha_{0}&=&\frac{4\,\rho\,z} {a} \, f_{0}   \nonumber\\
\alpha_{1}&=&\frac{4\,\rho\,z} {a} \, f_{1}   \nonumber\\
\alpha_{2}&=&\frac{4\,\rho\,z} {a} \, f_{2}   \nonumber\\
\alpha_{3}&=&\frac{4\,\rho^{2}}{a}\:\,f_{3}   \nonumber\\
\alpha_{4}&=&\frac{4\,\rho^{2}}{a}\:\,f_{4}   \nonumber\\
\beta_{1 }&=&\frac{2\,\rho\,z} {a} \, h_{1}   \nonumber\\
\beta_{2} &=&h_{2}                            \hspace{8.2cm}
\label{eq:ansatzalpha}
\end{eqnarray}
and
\begin{eqnarray*}
a &\equiv& \rho^{2} + z^{2} + r_{a}^{2}       \: ,\hspace{6.0cm}
\end{eqnarray*}
\noindent where $r_{a}$ is an arbitrary scale parameter.
The axial functions $f_{\mu}=f_{\mu}(\rho,z)$
and $h_{\nu}=h_{\nu}(\rho,z)$, with $\mu=0,1,2,3,4$ and $\nu=1,2$,
are non-singular and have reflection symmetry
\newcommand {\refansatzWa} {\ref{eq:ansatzW}, \ref{eq:ansatzalpha}}
\begin{eqnarray}
f_{\mu}(\rho,z)        &=& f_{\mu}(\rho,-z)  \nonumber \\
h_{\nu}(\rho,z)        &=& h_{\nu}(\rho,-z)  \: ,
\label{eq:frsymm}
\end{eqnarray}
Neumann \bcs ~on the z--axis
\begin{eqnarray}
\partial_{\rho} f_{\mu}(0,z) &=& 0           \nonumber\\
\partial_{\rho} h_{\nu}(0,z) &=& 0 \: ,
\label{eq:Neumannbcs}
\end{eqnarray}
and Dirichlet \bcs ~at infinity
\beq
\lim_{|x| \rightarrow \infty}
\left( \begin{array}{c}
       f_{0}\\ f_{1}\\ f_{2}\\ f_{3}\\ f_{4} \\ h_{1}\\ h_{2}
       \end{array}
\right) =
\left( \begin{array}{c}
       1\\-1 \\- \cos 2\theta\\ 1 + \cos 2\theta \\ 0 \\ 1\\- \cos 2\theta
       \end{array}
\right) \; .
\label{eq:infinitybcs}
\eeq
The additional condition
\beq
f_{1}(0,z) = f_{2}(0,z)
\label{eq:f1f2bcs}
\eeq
ensures having regular gauge field configurations on the z--axis.
The ansatz (\refansatzWa) is self-consistent, as will be explained in sect. 3.
It remains to solve for the axial functions $f_{\mu}$ and $h_{\nu}$ of
the ansatz. This will be done numerically and the results will be presented
in sect. 4.

The ansatz (\refansatzWa), restricted to two independent axial functions
$f(\rho,z)$ and $h(\rho,z)$, reproduces, up to a global gauge transformation,
the field configurations at the top of the \ncs ~as constructed in \cite{K90}.
Specifically, the restricted  axial functions $f_{\mu}$ and $h_{\nu}$ are
\begin{eqnarray}
 f_{0} & = &  \frac{(d^{2}/4+\rho^{2}+z^{2})\,a}{b} \, f       \nonumber\\
 f_{1} & = &  \frac{(d^{2}/4-\rho^{2}-z^{2})\,a}{b} \, f       \nonumber\\
 f_{2} & = &  \frac{(d^{2}/4+\rho^{2}-z^{2})\,a}{b} \, f       \nonumber\\
 f_{3} & = &  \frac{2\,  z^{2} \, a            }{b} \, f       \nonumber\\
 f_{4} & = &  \, 0                                             \nonumber\\
 h_{1} & = &  \frac{a                          }{\sqrt{b}}\: h \nonumber\\
 h_{2} & = &  \frac{d^{2}/4+\rho^{2}-z^{2}     }{\sqrt{b}}\: h \; ,
\end{eqnarray}
with
\[
b \equiv \left( \rho^{2} + (z-d/2)^{2} \right)
         \left( \rho^{2} + (z+d/2)^{2} \right)
\]
and the \bcs
\begin{eqnarray*}
f(0,\pm \,d/2)                          &=& 0 \\
h(0,\pm \,d/2)                          &=& 0 \\
\lim_{|x| \rightarrow \infty} f,\, h    &=& 1 \; .
\end{eqnarray*}
The parameter $d$ of these configurations sets
the distance between the two points on the z--axis
where the Higgs fields vanishes, as will be discussed in the next section.
The present ansatz (\refansatzWa) is thus the axisymmetric
generalization of a particular maximum configuration of a \ncs,
with the number of independent axial functions increased from two to seven.
The ansatz (\ref{eq:ansatzW}) has indeed the most general axisymmetric
form, as discussed by Manton \cite{M78}, whose
notation we follow, up to some trivial modifications.
The essential dynamics resides in the  structure
(\ref{eq:ansatzalpha}--\ref{eq:f1f2bcs}) for the axial functions
$\alpha_{\mu}$ and $\beta_{\nu}\,$, see sect. 3 below.

At this point we can make three general remarks.
First, we note that the behaviour at infinity of the $SU(2)$ gauge and Higgs
fields of \Sstar, in the radial gauge, is essentially the
same as for the well-known \sph ~S \cite{DHN74,KM84}, except for the
dependence on the polar angle $\theta$, which runs twice as fast
(i.e. $2\,\theta$ replacing $\theta$) \cite{K85}.
This suggests the interpretation of \Sstar ~as a bound state of two \sphs ~S,
or, more precisely, a \sph ~S and an anti-\sph ~$\bar{\rm S}$.
Secondly, we remark that the $U(1)$ field of our ansatz is consistent with
having two magnetic dipoles aligned along the z--axis,
with the behaviour towards infinity $ (B_{1},B_{2},B_{3}) \propto
\tan^{2}\thetaw \:r^{-3}\,(x_{2},-x_{1},0)\,$.
This would be a consequence of the interpretation of \Sstar ~as
an S--$\bar{\rm S}$ bound state and
the fact that the \sph ~S itself has a magnetic dipole field  \cite{KM84}.
Thirdly, we observe that the fields (\refansatzWa) of \Sstar ~respect
parity reflection symmetry, which is not
the case for the configurations of the \ncs ~in general
(the same holds for the \sph ~S and the corresponding  \ncl).
In fact, there are only two configurations on the minimal \ncs ~\cite{K90}
that have parity reflection symmetry, namely the vacuum and the \sph ~\Sstar.
This completes, for the moment, our discussion of the ansatz.

\section{Energy and field equations}

It is a straightforward, but tedious, exercise to insert the ansatz
(\refansatzWa) into the energy functional (\ref{eq:E}).
In terms of dimensionless distances (defined by $\tilde{x}_{m} \equiv
x_{m} M_{W}$, and dropping the tildes) we find for the energy of the
ansatz
\beq
E = \frac{4 \pi v}{g}
    \int_{0}^{\infty}dz \; \int_{0}^{\infty} d\rho\,\rho
    \;\; e(\rho,z) \; ,
\label{eq:Eansatz}
\eeq
where the dimensionless energydensity $e(\rho,z)$ is
an even function of $z$. Specifically, the energydensity is given by
\beq
e = e_{\rm WKIN}+e_{\rm BKIN}+e_{\rm HKIN}+e_{\rm HPOT}+e_{\rm GFIX} \; ,
\label{eq:eansatz}
\eeq
with
\newcommand {\tanw} {\tan \theta_{w}}
\newcommand {\tansqw} {\tan^{2} \theta_{w}}
\begin{eqnarray}
e_{\rm WKIN}
            &=&+ \; 4 \,          \left[ \,                  % w23v
          \partial_{\rho}\left(\frac{\rho\, f_{0}}{a}\right)
       -  \partial_{z}   \left(\frac{z\,    f_{1}}{a}\right)
                                        \, \right]^{2}  \nonumber\\
             & &+ \; 4 \,         \left[ \,                  % - w12u
          \partial_{\rho}\left(\frac{z\, f_{2}}{a}\right)
       +  4\,\frac{\rho\,z}{a^{2}}  \,   f_{1}\,f_{3}
       -  \frac{z}{\rho}\,\frac{1}{a}\, (f_{1}-f_{2})
                                             \, \right]^{2}  \nonumber\\
             & &+ \; 4 \,         \left[ \,                 % w23u
          \partial_{z   }\left(\frac{z\, f_{2}}{a}\right)
       + 4\,\frac{\rho^{2}}{a^{2}}\,f_{0}\,f_{3}
       -    \frac{1}{a} \,f_{0}
                                             \, \right]^{2}  \nonumber\\
             & &+ \; 4 \,         \left[ \,                 % - w12w
         \partial_{\rho}\left(\frac{\rho\,f_{3}}{a}\right)
       - 4\,\frac{z^{2}}{a^{2}}\,f_{1}\,f_{2}
       +    \frac{1}{a} \,f_{3}
                                             \, \right]^{2}  \nonumber\\
             & &+ \; 4 \,         \left[ \,                 % w23w
         \partial_{z}\left(\frac{\rho\, f_{3}}{a}\right)
       - 4\,\frac{\rho\,z}{a^{2}}\,f_{0}\,f_{2}
                                             \, \right]^{2}  \\
\label{eq:eWKIN}
&&\nonumber\\
e_{\rm BKIN} &=&+ \;      \tansqw \,  \left( \:         % - y121 and y23c
            4\,\left[ \,   \partial_{\rho}\left(\frac{\rho\,f_{4}}{a}\right)
                   +    \frac{1}{a} \,f_{4}
            \,     \right]^{2}
            \:+\: 4\,  \left[ \,
            \partial_{z}\left(\frac{\rho\, f_{4}}{a}\right)
            \,     \right]^{2}      \: \right)              \\
\label{eq:eBKIN}
&&\nonumber\\
e_{\rm HKIN} &=&+ \, \left[ \,                             % dphi2u
         \partial_{\rho}\left(\frac{2\,\rho\, z\, h_{1}}{a}\right)
       - 2\, \frac{z}{a} \,f_{1}\,h_{2}
                             \, \right]^{2}
                + \, \left[ \,
         \partial_{\rho} h_{2}                                   % - dphi2w
       + 2\,\frac{z}{a}\,f_{1}\,\frac{2\, \rho\,z\,h_{1}}{a}
                             \, \right]^{2}  \nonumber\\
             & &+ \, \left[ \,                             % dphi3u
         \partial_{z}\left(\frac{2\,\rho\, z\, h_{1}}{a}\right)
       - 2\, \frac{\rho}{a} \,f_{0}\,h_{2}
                             \, \right]^{2}
                + \, \left[ \,
         \partial_{z} h_{2}                                   % - dphi3w
       + 2\,\frac{\rho}{a}\,f_{0}\,\frac{2\,\rho\,z\,h_{1}}{a}
                             \, \right]^{2}  \nonumber\\
             & &+ \, \left[ \,                       % - dphy2v
         2\,\frac{\rho}{a}\,(f_{3}+f_{4}\, \tansqw )
                              \,\frac{2\,\rho\,z\, h_{1}}{a}
       - 2\,\frac{z}{a} \, ( h_{1} - f_{2}\,h_{2})
                             \, \right]^{2}  \nonumber\\
             & &+ \, \left[ \,                       % - dphy21
         2\,\frac{\rho}{a}\,(f_{3}-f_{4}\, \tansqw )\, h_{2}
       - 2\,\frac{z}{a} \, f_{2}\, \,\frac{2\,\rho\,z\, h_{1}}{a}
                             \, \right]^{2}           \\
\label{eq:eHKIN}
&&\nonumber\\
e_{\rm HPOT} &=&+ \, \, 2 \,\frac{\lambda}{g^{2}}   \left[ \,
           \frac{4\,\rho^{2}\, z^{2} \,h_{1}^{2}}{a^{2}}              % h4root
       +   h_{2}^{2}
       - 1
                                                \, \right]^{2}  \: ,
\label{eq:eHPOT}
\end{eqnarray}
and, for the moment, $e_{\rm GFIX}$ is set to zero.
The energydensity  is finite everywhere and vanishes at infinity,
for the appropriate behaviour of the axial functions.

We turn now to the classical field equations (\ref{eq:fieldeqs}).
Again, it is a straightforward exercise to insert the ansatz (\refansatzWa).
We find that the field equations reduce to seven partial differential equations
for the seven functions $f_{\mu}$ and $h_{\nu}$, which
are identical
\footnote{
This is a manifestation of the so-called
principle of symmetric criticality \cite{P79}, which states that,
under certain conditions, it suffices to consider variations that
respect the symmetry of the ansatz.
          }
to the variational equations from the ansatz energy
(\ref{eq:Eansatz}, \ref{eq:eansatz}). In short, our ansatz is self-consistent.

The solution of these partial differential equations for
$f_{\mu}(\rho,z)$ and $h_{\nu}(\rho,z)$
depends on their boundary conditions
(\ref{eq:Neumannbcs}--\ref{eq:f1f2bcs}) on the half--plane.
The \bcs ~for $h_{2}$, in particular, result in a curve $h_{2}=0$ in
the $\rho,z$--plane that comes in from infinity at an angle $\theta = \pi/4$
and then hits, by necessity, either the $z$--axis or $\rho$--axis.
With $\beta_{1}=0$ automatically on both axes, these points
$\beta_{2} = h_{2}= 0$ have vanishing Higgs field $\Phi$ altogether.
The two simplest possibilities correspond to having $\Phi=0$ at two points
on the z--axis or on a ring in the $z=0$ plane (these two points
or the ring may collapse to a single point at the orgin).
Wether or not such a molecule-like
or vortex-like configuration is realized follows from the
solution of the field equations. Of course, the ansatz used
should be  sufficiently general, in order to allow for both possibilities.
Before we solve the field equations, there is one technical point
that must be clarified.

The ansatz (\ref{eq:ansatzW}) still has a residual $U(1)$ gauge symmetry
\cite{M78}, which is generated by the $SU(2)$ transformation matrix
$\exp [\omega(\rho,z) \: \matv]$.
Under these transformations $\alphatz\equiv \alpha_{0}/z$ and
$\alphatr\equiv \alpha_{1}/\rho$
behave as 2-dimensional $U(1)$ gauge fields and the four other functions
combine into two complex scalars ($\alpha_{4}$ is invariant, of course).
In order to eliminate this extra degree of freedom we choose the Lorentz gauge
for our effective 2-dimensional euclidean gauge theory
\beq
     \partial_{z} \: \alphatz \:+\: \partial_{\rho} \: \alphatr \: = 0 \: ,
\label{eq:Lgauge}
\eeq
which has also been used (under a different name) for the
sphaleron S \cite{KKB92}.
%
% note factor 16 taken out in GFIX
%
This condition can be implemented by adding a gauge fixing term
to the energydensity (\ref{eq:eansatz})
\beq
e_{\rm GFIX} = +\, \xi              \, \left[ \,
          \partial_{z   }\left(\frac{\rho\, f_{0}}{a}\right)
       +  \partial_{\rho}\left(\frac{z\,    f_{1}}{a}\right)
                                        \, \right]^{2} \; ,
\label{eq:eGFIX}
\eeq
with $\xi$ an arbitrary parameter.

To summarize, we have shown that the ansatz (\refansatzWa)
solves the field equations, provided the axial functions
stationarize the energy integral (\ref{eq:Eansatz}).
Concretely, we have to solve the variational
equations for $f_{\mu}$ and $h_{\nu}$, with $\mu = 0,\,1,\,2,\,3,\,4$ and
$\nu=1,\,2$, that result from variations $\delta f_{\mu}$ and $\delta h_{\nu}$
in (\ref{eq:Eansatz}, \ref{eq:eansatz}), with the gauge fixing term
(\ref{eq:eGFIX}) included.
It does not seem possible, however, to obtain an analytical solution for these
functions and we have to resort to numerical methods.

\section{Numerical solution}

In this section we solve numerically the variational equations from the
energy integral (\ref{eq:Eansatz}, \ref{eq:eansatz}).
These variational equations consist of seven coupled non-linear
partial differential equations (PDEs).
Results are obtained for the case of approximately vanishing Higgs mass
$\lambda/g^{2}=1/800$ (or $M_{H}/M_{W} = 1/10$)
and weak mixing angle $\thetaw =0$. We also have
some preliminary results for the difficult, but more realistic, case
$\lambda/g^{2}=1/8$ (or $M_{H}/M_{W} = 1$)
and $\thetaw =\pi/6$.
As to the numerical method, the variational equations are solved
by numerical relaxation of the discretized energy
(\ref{eq:Eansatz}, \ref{eq:eansatz}) and
further details can be found in Appendix A of \cite{K93}.

For the case $\lambda/g^{2}=1/800$ and $\thetaw =0$,
we solve the variational equations for the axial functions
$f_{0}$, $f_{1}$, $f_{2}$, $f_{3}$, $h_{1}$ and $h_{2}$ ($f_{4}=0$).
The general behaviour of these functions is  shown in fig. 1.
In particular, we see the $h_{2}=0$ curve coming in from infinity
at a polar angle $\theta=\pi/4$ and hitting the z--axis just below
$z=2 \, M_{W}^{-1}$. As discussed in the previous section, this means that
the solution has two points $z=\pm\, \dstar /2$ on the symmetry axis
where the Higgs field $\Phi$ vanishes.
The resulting estimates for the internal core distance and energy of
the \sph ~\Sstar ~are
\begin{eqnarray}
\dstar &\sim& (4 \pm 2) \; M_{W}^{-1}             \nonumber\\
E_{{\rm S}^{\star }} &\sim& (1.91 \pm 0.02)\; \ES \: ,
\label{eq:dstar00}
\end{eqnarray}
where  $\ES = 1.57 \; 4\pi v/g$ is the energy of the \sph ~S.
The corresponding energydensity distribution (fig. 2) resembles that of a
diatomic molecule, but the exact solution of the PDEs may very well be
tighter than the one obtained numerically.
Note that our earlier approximation of \Sstar ~from the
\ncs ~results in \cite{K93} gave essentially the same energy, but a
larger core distance ($d_{\rm NCS} \sim 7 \: M_{W}^{-1}$).

For the case $\lambda/g^{2}=1/8$ and $\thetaw =\pi/6$,
we solve the variational equations for all seven axial functions
$f_{\mu}$ and $h_{\nu}$. The preliminary results are
$\dstar \sim (6 \pm 3) \; M_{W}^{-1}$ and
$E_{{\rm S}^{\star }} \sim (1.99 \pm 0.02)\; \ES$.
Clearly, the numerical accuracy has to be improved to obtain reliable
numbers. On physical grounds, though,  we expect an energy value below
$2 \ES$ and a finite core distance, as will be explained in the next section.

To summarize, the solution of the field equations for our ansatz
appears to be a rather difficult numerical problem.
Still, we have managed to obtain a first estimate of the solution,
in particular for the case $\lambda / g^{2} \sim  0$ and $\thetaw =0$.

\section{Discussion}

We have constructed in this article a self-consistent ansatz for a new
\sph ~\Sstar ~in the \ewsmodel. The axial functions of the solution
can only be determined numerically.  Wether or not the core distance $\dstar$
between the points of vanishing Higgs field
remains finite is a dynamical question.
In fact, the \sph ~\Sstar ~may be considered as a bound state
of a \sph ~S and an anti-\sph ~$\bar{\rm S}$,
each with vanishing Higgs field at the center.
Physically, we have for small values of the distance $d\,$ between
S and $\bar{\rm S}$ repulsion from the Yang-Mills fields
and for large values of $d\,$ attraction from
either the Higgs scalar (if $M_{H} < M_{W}$)
or from the parallel magnetic dipoles
of the photon field (provided $\thetaw > 0$).
Hence we expect, for generic values of $M_{H}/M_{W}$ and $\thetaw$,
a finite core distance $\dstar$ of the \sph ~\Sstar, or, in other words,
a localized solution.

In this paper we have employed a simple procedure to arrive at a new
self-consistent ansatz,
namely by generalizing the maximum configuration of a \ncs, while
keeping the symmetries of the configurations of the \ncs.
It is important to note that our maximum configuration is distinguished
by an additional (discrete) symmetry.  This construction method,
which is obviously inspired by equivariant Morse theory \cite{SSU89},
can probably also be used to discover other classical solutions.
In particular, it should be possible to obtain a
\scansatz ~for a new constrained instanton \Istar, for which we have found a
\ncl ~of 4-dimensional euclidean configurations \cite{K93}.
There is once more a rotational symmetry, which must be maintained when
the number of axial functions is increased for the maximum configuration
of the \ncl.
\footnote{
{}~In the ansatz of \cite{K93}, at the top ($\omega=0$) of the \ncl,
we first set the additional functions
$g_{\pm}=1$, then replace the functions $f(r,\tau)$, $h(r,\tau)$
by the more general ones $f(\rho,z,\tau)$, $h(\rho,z,\tau)$,
and finally  multiply their number
to $f_{\mu}(\rho,z,\tau)$, $h_{\nu}(\rho,z,\tau)$.
          }
The resulting field equations will have to be solved numerically.
Clearly, this will be quite demanding technically, but, in principle,
we see no obstacle to the explicit construction of \Istar.
Note that, loosely speaking, the \sph ~\Sstar ~corresponds to a constant time
slice of the instanton \Istar, as explained in \cite{K93}.

This brings us back to the \sph ~\Sstar ~and its possible physical relevance.
The crucial observation seems to be
that \Sstar ~sets the minimum height of the energy barrier for Witten's
global $SU(2)$ anomaly, in the hamiltonian formulation \cite{W82}.
This can be explained as follows.
In hamiltonian $SU(2)$ gauge theory there
exist \ncls ~of 3-dimensional gauge transformations (based on non-trivial
maps ${\rm S}_{1}\times{\rm S}_{3}\rightarrow SU(2) \sim {\rm S}_{3}$)
that may interfere with the implementation of Gauss' law,
thereby giving rise to the anomaly.
Consider now a non-contractible orbit of an arbitrary  vacuum
configuration (energy $E=0$). This loop can be made contractible, simply
by lifting it out of the vacuum and pulling it over an energy barrier (fig.
3a).
The \sph ~\Sstar ~sits at the top of this
barrier and $E_{\rm barrier} = E_{{\rm S}^{\star }}$.
Fixing the gauge completely (assuming the anomalies to cancel)
results in  a \ncs ~of 3-dimensional
configurations (fig. 3b), with the unique vacuum V as its lowest point ($E=0$)
and the \sph ~\Sstar ~as its highest ($E=E_{{\rm S}^{\star }}$).
This \ncs ~has spectral flow of the eigenvalues of the Dirac operator
\cite{W82,NAG85}.
Most likely, a pair of eigenvalues crosses through zero precisely at
the top of the barrier, and the \sph ~\Sstar ~may be expected to have
fermion zero modes.

The role of \Sstar ~as the barrier height for the global $SU(2)$ anomaly
is similar to that of the standard \sph ~S for the chiral $U(1)$ anomaly.
Both anomalies arise from the spectral flow of Dirac eigenvalues,
over \ncss ~for the case of \Sstar ~and over \ncls ~for the case of S.
The physical consequences of these two \sphs ~are, however, entirely
different, since in the \sm ~only the chiral anomaly is coupled to
a global quantum number ($B+L$). Moreover, the \sm ~has no $SU(2)$
anomaly to begin with, because the number of left-handed fermion doublets
is even. For this reason we expect the new \sph ~\Sstar ~to play a role in the
consistency and dynamics of the electroweak theory, rather than produce
qualitatively new physical effects.
Indeed, the most important application of the \ew ~\sph ~\Sstar, or more
appropriately its related constrained instanton \Istar ~\cite{K93,K92},
may turn out to be for the asymptotics of perturbation theory \cite{GZJ90}
and multiparticle production \cite{AG87,R90}.
These fundamental problems deserve further study and
perhaps some progress can be made with our new classical solutions.

\vspace{2\baselineskip}
\section*{Acknowledgements}
We thank the experimental colleagues at NIKHEF-H for access to their Apollo
workstations and the staff of the Computer Group for technical assistance.

\newpage
\section*{Figure captions }
\par
{\bf Fig. 1 :} Equidistant contours for the axial functions ($f_{4}=0$)
in the ansatz (\refansatzWa) of the \sph ~\Sstar.
These functions give the numerical solution of the field equations,
evaluated for the ansatz.
The scale parameter of the ansatz is $r_{a}= M_{W}^{-1}$ and
the coordinates $\rho$ and $z$ are also in units of $M_{W}^{-1}$.
The coupling constants of the theory are $\lambda/g^{2}=1/800$
and  $\thetaw =0$.
\vspace{1\baselineskip}
\newline
{\bf Fig. 2:} Energydensity $e(\rho,z)$ (with arbitrary normalization)
of the \sph ~\Sstar, corresponding to the axial functions of fig. 1.
The coordinates $\rho$ and $z$ are in units of $M_{W}^{-1}$ and
the  complete configuration is obtained by reflection of $z$.
\vspace{1\baselineskip}
\newline
{\bf Fig. 3 :} {\bf (a)} Sketch of the energy surface over \cs.
A \ncl ~in the vacuum, with fixed basepoint,
becomes trivial by sliding over an energy barrier.
The heigth of this energy barrier is given by the \sph ~\Sstar.
A more realistic picture of \cs ~has antipodal points on the disk shown
identified and the vacuum circle replaced by the $SO(3)$ manifold,
in order to exhibit the cyclicity of the first homotopy group. \newline
{\bf (b)} \Ncs ~in configuration space, with the gauge completely fixed.

\end{document}